\documentstyle[times,pramana,epsfig,floats]{ias}

\newcommand{\dslash}{\partial \hskip -0.5em /}
\newcommand{\vslash}{v \hskip -0.5em /}
\newcommand{\bD}{{\bf D}}

\newcommand{\xipl}{\vec{\xi\,}\hskip-0.6mm
+\hskip-0.6mm\lambda\hat{e}_3}
\newcommand{\ximl}{\vec{\xi}\hskip-0.6mm
-\hskip-0.6mm\lambda\hat{e}_3}
\newcommand{\zr}[1]{\mbox{\hspace*{#1em}}}
\newcommand{\ID}{\mbox{{\sf 1}\zr{-0.16}\rule{0.04em}{1.55ex}\zr{0.1}}}

\begin{document}
\title{Chiral Quark Model\thanks{Talk presented at the 
workshop QCD 2002, IIT Kanpur, Nov. 2002.}}

\author{H Weigel\thanks{Heisenberg--Fellow}}
\address{Institute for Theoretical Physics, T\"ubingen University,
Auf der Morgenstelle 14, D--72076 T\"ubingen, Germany}

\abstract{In this talk I review studies of hadron properties in
bosonized chiral quark models for the quark flavor dynamics. 
Mesons are constructed from Bethe--Salpeter equations and 
baryons emerge as chiral solitons. Such models require regularization
and I show that the two--fold Pauli--Villars regularization scheme
not only fully regularizes the effective action but also
leads the scaling laws for structure functions. For the nucleon
structure functions the present approach serves to determine
the regularization prescription for structure functions whose
leading moments are not given by matrix elements of local
operators. Some numerical results are presented for the spin
structure functions. 
}

\maketitle

\section{Introduction}
In this talk I review investigations of hadron properties in the 
Nambu--Jona--Lasino (NJL) model\cite{Na61}. This is a particularly 
simple model for the quark flavor interactions with the important 
feature that the quarks can be integrated out in favor of meson
fields\cite{Eb86}. The resulting effective action for these 
mesons possesses soliton solutions\cite{Al96}. According to the 
large--$N_C$ picture\cite{tH74} of Quantum--Chromo--Dynamics (QCD) 
these solutions are interpreted as baryons. 

The construction of hadron wave--functions is not possible 
in QCD. This represents a main obstacle for the computation of 
hadron properties from first principles. As the NJL model adopts the 
symmetry properties of QCD, the current operators in the model 
correspond to those of QCD. As a consequence, matrix elements of 
the current operators as computed in the model are sensible and
their comparison with experimental data is meaningful. In particular, 
it is interesting to analyze the hadronic tensor that parameterizes 
the deep--inelastic--scattering (DIS) and confront 
the model predictions with empirical data. This picture has led 
to interesting studies of hadron structure functions in bosonized
chiral quark models. Here I will present the results of
refs.\cite{We96a,We97,We99}. These studies build up a 
consistent approach by computing the hadronic tensor (or equivalently
the forward virtual Compton amplitude) from the gauged meson action.
For the nucleon structure functions similar studies have been reported 
in refs.\cite{Di96,Wa98,Wa00}. There no attempt to compute the 
structure functions from the gauged action was made but rather 
it was assumed that the model predictions for the constituent 
quark distributions can be identified with QCD 
quark distributions. I refer to those articles for a more 
expatiated presentation of numerical results. In addition, I 
refer to the review articles~\cite{Al96} for comprehensive discussions
of model predictions for static baryon properties such as magnetic
moments, axial charges or the hyperon spectrum.

This talk is organized as follows. In Section~2 I introduce 
the NJL model as an effective meson theory and utilize pion
properties to determine the model parameters. Section~3 describes
the subtleties for extracting the structure functions that
arise in this model from regularization. The pion structure 
function is considered as an example. In Section~4 I
review the construction of baryon states in the soliton
picture. The following Section sketches the computation of
nucleon matrix elements of the hadronic tensor and the 
extraction of the structure functions in the Bjorken limit.
Finally in Section~6 I present some numerical results
for the spin structure functions $g_1$ and $g_2$ and compare them 
to experimental data by means of the transformation to the infinite 
momentum frame and subsequent DGLAP evolution. Section~7 serves as a 
short summary.
 
\section{The NJL Model for Chiral Dynamics}

The NJL model is a quark model with a chirally invariant quartic 
quark interaction. Bosonization is achieved semiclassically by
introducing effective meson fields for the
quark bilinears in that interaction. Then the quark fields are
integrated out by functional methods. This yields an effective 
action for meson degrees of freedom,
\begin{equation}
{\mathcal A} [S,P]=-iN_C{\rm Tr}_{\textstyle\Lambda}{\rm log}\, 
\left[i\dslash-\left(S+i\gamma_5P\right)\right]
-\frac{1}{4G}\int d^4x\, {\rm tr}\, {\mathcal V}(S,P)\, .
\label{bosact}
\end{equation}
Here ${\mathcal V}$ is a local potential for the effective scalar and 
pseudoscalar fields $S$ and $P$, respectively, that are matrices in 
flavor space. In the NJL model the potential reads
${\mathcal V}=S^2+P^2+2{\hat m}_0S$ with $\hat{m}_0$ being the
current quark mass matrix. Since the interaction is mediated by 
flavor degrees of freedom, the number of colors, $N_C$, is merely 
a multiplicative quantity. The functional trace~(${\rm Tr}$) 
originates from integrating out the quarks and induces a non--local
interaction for $S$ and~$P$. For simplicity I will only consider the 
isospin limit for up (u) and down (d) quarks: $m_{0,u}=m_{0,d}=m_0$.

A major concern in regularizing the functional (\ref{bosact}), as 
indicated by the cut--off $\Lambda$, is to maintain the chiral 
anomaly. This is achieved by splitting this functional into 
$\gamma_5$--even and odd pieces and only regulate the 
former,
\begin{eqnarray}
&&{\rm Tr}_\Lambda {\rm log}\,
\left[i\dslash-\left(S+i\gamma_5P\right)\right]
=-i\frac{N_C}{2} \sum_{n=0}^2 c_n {\rm Tr}\, {\rm log}
\left[- \bD \bD_5 +\Lambda_n^2-i\epsilon\right]
\hspace{0.4cm} \nonumber \\ &&\hspace{4.5cm}
-i\frac{N_C}{2}
{\rm Tr}\, {\rm log}
\left[-\bD \left(\bD_5\right)^{-1}-i\epsilon\right]\, ,
\label{PVreg} \\
{\rm with}\,&&\qquad
i \bD = i\dslash - \left(S+i\gamma_5P\right) 
\quad {\rm and} \quad
i \bD_5 = - i\dslash - \left(S-i\gamma_5P\right)\, .
\label{defd}
\end{eqnarray}
The double Pauli--Villars regularization renders the functional
(\ref{bosact}) finite with 
$c_0=1,\,\, \Lambda_0=0 ,\, \sum_{n=0}^2c_n=0\,.$
The $\gamma_5$--odd piece is
conditionally finite and not regularizing it, reproduces the chiral 
anomaly properly.  For sufficiently large $G$ one 
obtains the VEV, $\langle S\rangle\equiv m\ID$ that parameterizes 
the dynamical chiral symmetry breaking from the gap--equation, 
\begin{equation}
\frac{1}{2G}\left(m-m_0\right)
=4iN_C m\sum_{n=0}^2c_n
\int\frac{d^4k}{(2\pi)^4}
\left[k^2-m^2-\Lambda_n^2+i\epsilon\right]^{-1}\, .
\label{gap}
\end{equation}
Substituting $S=\langle S\rangle=m\ID$ in 
eq.~(\ref{bosact}) shows that $m$ plays the role of a 
mass and is therefore called the constituent quark mass.

In the next step I utilize pion properties to fix the model
parameters and introduce the isovector pion field ${\vec\pi}$ via
\begin{equation}
S+iP\gamma_5=m\, \left(U\right)^{\gamma_5} = m\, {\rm exp}
\left(i \frac{g}{m}\gamma_5\,{\vec\pi} \cdot {\vec\tau} \right)\, .
\label{SandP}
\end{equation}
Sandwiching the axial current between the 
vacuum and a single pion state yields the pion decay constant 
$f_\pi=93{\rm MeV}$ in terms of the polarization function 
$\Pi(q^2,x)$,
\begin{eqnarray}
f_\pi&=&4N_Cmg\int_0^1\, dx\, \Pi(m_\pi^2,x)
\nonumber \\ 
\Pi(q^2,x)&=&\sum_{n=0}^2 c_n\, \frac{d^4k}{(2\pi)^4i}\,
\left[k^2+x(1-x)m_\pi^2-m^2-\Lambda_n^2+i\epsilon\right]^{-2}\, ,
\label{fpi}
\end{eqnarray}
where $m_\pi=138{\rm MeV}$ is the pion mass.
The Yukawa coupling constant, $g$, is determined by the requirement
that the pion propagator has unit residuum,
\begin{equation}
\frac{1}{g^2}=4N_C \frac{d}{dm_\pi^2}\int_0^1\, dx\,
\left[m^2_\pi \Pi(m_\pi^2,x)\right]\, .
\label{yukawa}
\end{equation}
In the chiral limit ($m_\pi=0$) this simplifies to $f_\pi=m/g$. 
Finally the current quark mass is fixed from the condition that
the pole of the pion propagator is exactly at the pion mass,
\begin{equation}
m_0=4\,N_C\,m\,G\, m_\pi^2\,
\int_0^1\, dx\, \Pi(m_\pi^2,x)\, .
\label{mpi}
\end{equation}
It is also worthwhile to mention that expanding eqs.~(\ref{PVreg}) 
and (\ref{SandP}) to linear and quadratic order  in $\vec{\pi}$ 
and $v_\mu$, respectively, yields the correct width for
the anomalous decay $\pi^0\to\gamma\gamma$. This is the direct 
consequence of not regularizing the $\gamma_5$--odd piece.

Before discussing nucleons as solitons of the bosonized 
action~(\ref{bosact}) and the respective structure functions
it will be illuminating to first consider DIS off pions.

\section{The Compton Tensor and Pion Structure Function}

DIS off hadrons is parameterized by the hadronic tensor 
$W^{\mu\nu}(p,q)$ where $q$ is the momentum transmitted from
the photon to the hadron with momentum $p$.

The tensor $W^{\mu\nu}(p,q)$ is obtained from the hadron matrix element 
of the current commutator by Fourier transformation and is parameterized 
in terms of form factors that multiply the allowed Lorentz structures. 
These form factors are obtained by pertinent projection of the hadronic
tensor.  Finally the structure functions are the leading
twist contributions of the form factors. These contributions
are obtained from computing $W^{\mu\nu}(p,q)$ in the Bjorken limit:
$Q^2=-q^2\to\infty$ with $x=Q^2/p\cdot q$ fixed. That is, subleading
contributions in $1/Q^2$ are omitted.

An essential feature of bosonized quark models is that the derivative 
term in (\ref{bosact}) is formally identical to that of a 
non--interacting (or asymptotically free) quark model. Hence the 
current operator is given as $J^\mu={\bar q}{\mathcal Q}\gamma^\mu q$, 
with ${\mathcal Q}$ a flavor matrix. Expectation values of currents are 
computed by introducing pertinent sources $v_\mu$ in eq.~(\ref{PVreg}) 
\begin{equation}
i\bD\, \longrightarrow \, i\bD +{\mathcal Q}\vslash
\qquad {\rm and} \qquad
i\bD_5\, \longrightarrow \, i\bD_5 -{\mathcal Q}\vslash
\label{source}
\end{equation}
and differentiating the gauged action~(\ref{bosact}) with
respect to $v_\mu$. In bosonized quark models it is 
convenient to start from the absorptive part of the forward 
virtual Compton amplitude\footnote{The momentum of the hadron is 
called $p$ and its spin eventually $s$.} 
\begin{equation}
\hspace{-1.7cm}
T^{\mu\nu}(p,q)=\int d^4\xi\, {\rm e}^{iq\cdot\xi}\,
\langle p,s|T\left(J^\mu(\xi) J^\nu(0)\right)|p,s\rangle
\,\,\,  ,  \,\,\,\, 
W^{\mu\nu}(p,q)=\frac{1}{2\pi} {\mathsf{Im}}\, [T^{\mu\nu}(p,q)]
\label{comp1}
\end{equation}
because the time--ordered product is straightforwardly obtained from 
\begin{equation}
\hspace{-1.0cm}
T\left(J^\mu(\xi) J^\nu(0)\right)=
\frac{\delta^2}{\delta v_\mu(\xi)\, \delta v_\nu(0)}\,
{\rm Tr}_\Lambda {\rm log}\,
\left[i\dslash-\left(S+i\gamma_5P\right)+{\mathcal Q}\,
\vslash\right]\Big|_{v_\mu=0}\, ,
\label{tprod}
\end{equation}
as defined from eq.~(\ref{PVreg}) with the substitution~(\ref{source}).

Pion--DIS is characterized by a single structure function,
$F(x)$. For its computation the pion matrix element in the Compton 
amplitude~(\ref{comp1}) must be specified.  For virtual pion--photon 
scattering it is obtained by expanding eqs.~(\ref{PVreg}) and 
(\ref{SandP}) to second order in both, 
${\vec\pi}$ and $v_\mu$. Due to the separation into $\bD$ and $\bD_5$ 
this calculation differs considerably from the simple evaluation of 
the `handbag' diagram. For example, isospin violating and 
dimension--five operators appear for the action~(\ref{PVreg}). 
Fortunately all isospin violating pieces cancel yielding 
\begin{eqnarray}
F(x)=\frac{5}{9} (4N_C g^2)\frac{d}{dm_\pi^2}
\left[m_\pi^2 \Pi(m_\pi^2,x)\right]\,,
\quad 0\le x\le1\,.
\label{pionf}
\end{eqnarray}
The same result is obtained by formal treatment of the divergent
handbag diagram and {\it ad hoc} regularization\cite{Fr94}.
The cancellation of the isospin violating pieces is a feature
of the Bjorken limit: insertions of the pion field on the propagator
carrying the infinitely large photon momentum can be safely ignored. 
Furthermore this propagator can be taken to be the one for 
non--interacting massless fermions. This implies that also the 
Pauli--Villars cut--offs can be omitted for this propagator. That, in 
turn, leads to the desired scaling behavior of the structure function
in this model and is a particular feature of the Pauli--Villars 
regularization. {\it A priori} it is not obvious for an arbitrary 
regularization scheme that terms of the form $Q^2/\Lambda_n^2$ drop 
out in the Bjorken limit. 

{}From eqs.~(\ref{yukawa}) and~(\ref{pionf}) it is obvious that 
$F(x)=5/9$ for $0\le x\le1$
in the chiral limit ($m_\pi=0$). It must be noted that this refers 
to the structure function at the (low) energy scale of the model. To 
compare with empirical data, that are at a higher energy scale,
the DGLAG program of perturbative QCD has to be applied to $F(x)$
to include the ${\rm ln}Q^2$ corrections. Such studies~\cite{Da01}
show good agreement with the experimental data for $F(x)$.

\section{The Nucleon as a Chiral Soliton}

Solitons are a non--perturbative 
stationary configurations of the meson fields. To 
determine that configuration for the meson theory~(\ref{bosact})
I consider the hedgehog {\it ansatz}
$$
U_{\rm H}(\vec{r})={\rm exp}\left(i\vec{\tau}\cdot\hat{r}F(r)\right)
\quad {\rm and}\quad
\left(U_{\rm H}(\vec{r\,})\right)^{\gamma_5}=
{\rm exp}\left(i\gamma_5\vec{\tau}\cdot\hat{r}F(r)\right)
$$
for the pion field~(\ref{SandP}). The corresponding single 
particle Dirac Hamiltonian reads
\begin{equation}
h=\vec{\alpha}\cdot\vec{p} +\beta\, m\, \left[{\rm cos}F+
i\gamma_5 \vec{\tau}\cdot\hat{r}\, {\rm sin}F\right]\, .
\label{Dirac}
\end{equation}
Evaluating the action functional~(\ref{PVreg}) in the 
eigenbasis of $h$ gives the energy functional in 
terms of the eigenvalues, $\epsilon_\alpha$,~\cite{Do92} 
\begin{eqnarray}
E[F]&=&
\frac{N_C}{2}\left(1-{\rm sign}(\epsilon_{\rm V})\right)
\epsilon_{\rm V}
-\frac{N_C}{2}\sum_\alpha \sum_{n=0}^2 c_n 
\left\{\sqrt{\epsilon_\alpha^2+\Lambda_n^2}
-\sqrt{\epsilon_\alpha^{(0)2}+\Lambda_n^2}
\right\}
\nonumber \\ && \hspace{2cm}
+m_\pi^2f_\pi^2\int d^3r \, (1-{\rm cos}F)\, 
\label{etot}
\end{eqnarray}
for a baryon number one configuration. Here ${\rm V}$ denotes the 
unique quark level that is strongly bound by the soliton. Its 
explicit occupation takes care of the total fermion number and 
thus this level is referred to as the {\emph{valence quark}}. It
should not be confused with the valence quarks in the parton model.
Furthermore $\epsilon_\alpha^{(0)}$ are the eigenvalues of
$h^{(0)}=\vec{\alpha}\cdot\vec{p} +\beta\, m\,$.
The soliton profile $F(r)$ is then obtained from extremizing 
$E$ self--consistently~\cite{Al96}. 

States possessing good spin and isospin quantum numbers are 
generated by taking the zero--modes to be time dependent\cite{Ad83}
\begin{equation}
U(\vec{r\,},t)=
A(t)U_{\rm H}(\vec{r\,})A^{\dag}(t)\ ,
\label{collrot}
\end{equation}
which introduces the collective coordinates $A(t)\in SU(2)$. The 
action functional is expanded\cite{Re89} up to quadratic order in 
the angular velocities 
\begin{equation} 
i\vec{\tau}\cdot\vec{\Omega}=
2A^{\dag}(t)\dot A(t)\, .
\label{angvel}
\end{equation}
The coefficient of the quadratic\footnote{A liner term does 
not arise due to isospin symmetry.} term defines the moment
of inertia\footnote{Functional integrals are evaluated using
the eigenfunctions $\phi_\alpha$ of the Dirac Hamiltonian~(\ref{Dirac})
in the background of the chiral angle $F(r)$. Thus all quantities --
like the moment of inertia -- turn into functionals of $F(r)$.}, 
$\alpha^2[F]$. Nucleon states $|N\rangle$ are obtained 
by canonical quantization of the collective coordinates, $A(t)$. This
is analogous to quantizing a rigid rotator and allows to
compute matrix elements of operators in the space of the 
collective coordinates\cite{Ad83}:
\begin{equation}
\langle N |\textstyle{\frac{1}{2}}
{\rm tr}\left(\tau_a A^\dagger \tau_b A\right)|N\rangle=
-\textstyle{\frac{4}{3}}\langle N | I_a J_b|N\rangle
\quad {\rm and} \quad
\vec{\Omega}=-\vec{J\,}/\alpha^2[F]\,,
\label{nmatrix}
\end{equation}
where $I_a$ and $J_b$ denote isospin and spin, respectively.

For later use I note that the valence quark wave--function 
receives a first order cranking correction 
\begin{equation}
\Psi_{\rm V}(\vec{r\,},t)=
{\rm e}^{-i\epsilon_{\rm V}t}A(t)
\left\{\phi_{\rm V}(\vec{r\,})
+\frac{1}{2}\sum_{\mu\ne{\rm V}} \phi_\mu(\vec{r\,})
\frac{\langle \mu |\vec{\tau}\cdot\vec{\Omega}|{\rm V}\rangle}
{\epsilon_{\rm V}-\epsilon_\mu}\right\}\, ,
\label{valrot}
\end{equation}
where $\phi_\mu(\vec{r\,})$ are the eigenfunctions of $h$ in
eq.~(\ref{Dirac}).
The moment of inertia, $\alpha^2[F]$ is 
order $N_C$, thus, upon quantization~(\ref{nmatrix}), this rotational 
correction is subleading in $1/N_C$.

\section{Nucleon Structure Functions}

DIS off nucleons is described by four structure functions:
$F_1(x)$ and $F_2(x)$ are insensitive to the nucleon spin while
the polarized structure functions, $g_1(x)$ and $g_2(x)$, are associated 
with the components of the hadronic tensor that contain the nucleon spin. 

As argued in section~3, the quark propagator with the infinite 
photon momentum should be taken to be the one for free and massless
fermions. Thus, it is 
sufficient to differentiate (Here $\bD$ and $\bD_5$ are those
of eq~(\ref{defd}), {\it i.e.} with $v_\mu=0$.)
\begin{eqnarray}
&&
\hspace{-0.6cm}
\frac{N_C}{4i}\sum_{n=0}^2c_n
{\rm Tr}\,\left\{\left(-\bD\bD_5+\Lambda_n^2\right)^{-1}
\left[{\mathcal Q}^2\vslash\left(\dslash\right)^{-1}\vslash\bD_5
-\bD(\vslash\left(\dslash\right)^{-1}\vslash)_5
{\mathcal Q}^2\right]\right\}
\nonumber \\ &&
+\frac{N_C}{4i}
{\rm Tr}\,\left\{\left(-\bD\bD_5\right)^{-1}
\left[{\mathcal Q}^2\vslash\left(\dslash\right)^{-1}\vslash\bD_5
+\bD(\vslash\left(\dslash\right)^{-1}\vslash)_5
{\mathcal Q}^2\right]\right\}\, ,\,\,
\label{simple}
\end{eqnarray}
with respect to the photon field $v_\mu$. 
I have introduced the $(\ldots)_5$ description
$$
\gamma_\mu\gamma_\rho\gamma_\nu
=S_{\mu\rho\nu\sigma}\gamma^\sigma
-i\epsilon_{\mu\rho\nu\sigma}\gamma^\sigma\gamma^5
\, ,\quad 
(\gamma_\mu\gamma_\rho\gamma_\nu)_5
=S_{\mu\rho\nu\sigma}\gamma^\sigma+
i\epsilon_{\mu\rho\nu\sigma}\gamma^\sigma\gamma^5
$$
to account for the unconventional appearance of axial sources in 
$\bD_5$, {\it cf.} ref.\cite{We99}. Substituting eq.~(\ref{collrot})
for the meson fields that are contained in $\bD$ and $\bD_5$,
computing the functional trace up to subleading order 
in $1/N_C$ using a basis of quark states obtained from the 
Dirac Hamiltonian~(\ref{Dirac}), yields analytical results for 
the structure functions. I refer to ref.\cite{We99} for detailed 
formulas for other structure functions and the verification of the 
sum rules that relate integrals over the structure functions
to static nucleon properties. As an example I restrain myself to
list the contribution to $g_1(x)$ which is leading order in $1/N_C$:
\begin{eqnarray}
g_1(x)&=& \frac{M_NN_C}{36i}
\Big\langle N\Big| I_3 \Big| N\Big\rangle
\int \frac{d\omega}{2\pi} \sum_\alpha \int d^3\xi
\int \frac{d\lambda}{2\pi}\, {\rm e}^{iM_Nx\lambda}
\hspace{3cm}\nonumber \\ && \hspace{0cm}\times
\left(\sum_{n=0}^2\frac{c_n\left(\omega+\epsilon_\alpha\right)}
{\omega^2-\epsilon_\alpha^2-\Lambda_n^2+i\epsilon}\right)_{\rm P}
\Big[\phi^\dagger_\alpha(\vec{\xi\,})\tau_3
\left(1-\alpha_3\right)\gamma_5
\phi_\alpha(\xipl)
{\rm e}^{-i\omega\lambda}
\nonumber \\ && \hspace{4.0cm}
+\phi^\dagger_\alpha(\vec{\xi})\tau_3
\left(1-\alpha_3\right)\gamma_5
\phi_\alpha(\ximl)
{\rm e}^{i\omega\lambda}\Big]\, ,
\label{g1x}
\end{eqnarray}
where the subscript ($P$) indicates the pole term.

Before discussing numerical results I would like to mention the
unexpected result that the structure function entering the Gottfried sum rule 
is related to the $\gamma_5$--odd piece of the action and hence does not 
undergo regularization. This is surprising because in the parton model 
this structure function differs from the one associated with the Adler 
sum rule only by the sign of the anti--quark distribution. The latter 
structure function, however, gets regularized in the present model, 
in agreement with the quantization rule for the collective 
coordinates that correspond to the isospin operator that involves
the regularized moment of inertia, $\alpha^2$. 

\section{Numerical Results for Nucleon Structure Functions}

Unfortunately numerical results for the full structure functions 
in the double Pauli--Villars regularization scheme,
{\it i.e.} including the properly regularized vacuum piece are not yet 
available. However, in the Pauli--Villars regularization the axial 
charges are saturated to 95\% or even more by their valence quark 
(\ref{valrot}) contributions once the self--consistent soliton 
is substituted. This provides sufficient justification to consider 
the valence quark contribution to the polarized structure functions
as a reliable approximation since {\it e.g.} the zeroth moment
of the leading structure function~$g_1$ is nothing but the axial current
matrix element. This valence quark level is that of the chiral soliton 
model and, as already mentioned, its contributions to the structure 
functions should not be 
confused with valence quark distributions in parton models. In general,
it should be stressed that the present model calculation yields
structure functions, {\it i.e.} quantities that parameterize the
hadronic tensor, but not (anti)--quark distributions. The latter
would require the identification of model degrees of freedom 
with those in QCD. However, here only the symmetries (namely
the chiral symmetry) and thus the current operators in the hadronic
tensor are identified.

\begin{figure}[t]
~~\parbox[l]{5.2cm}{
\epsfig{figure=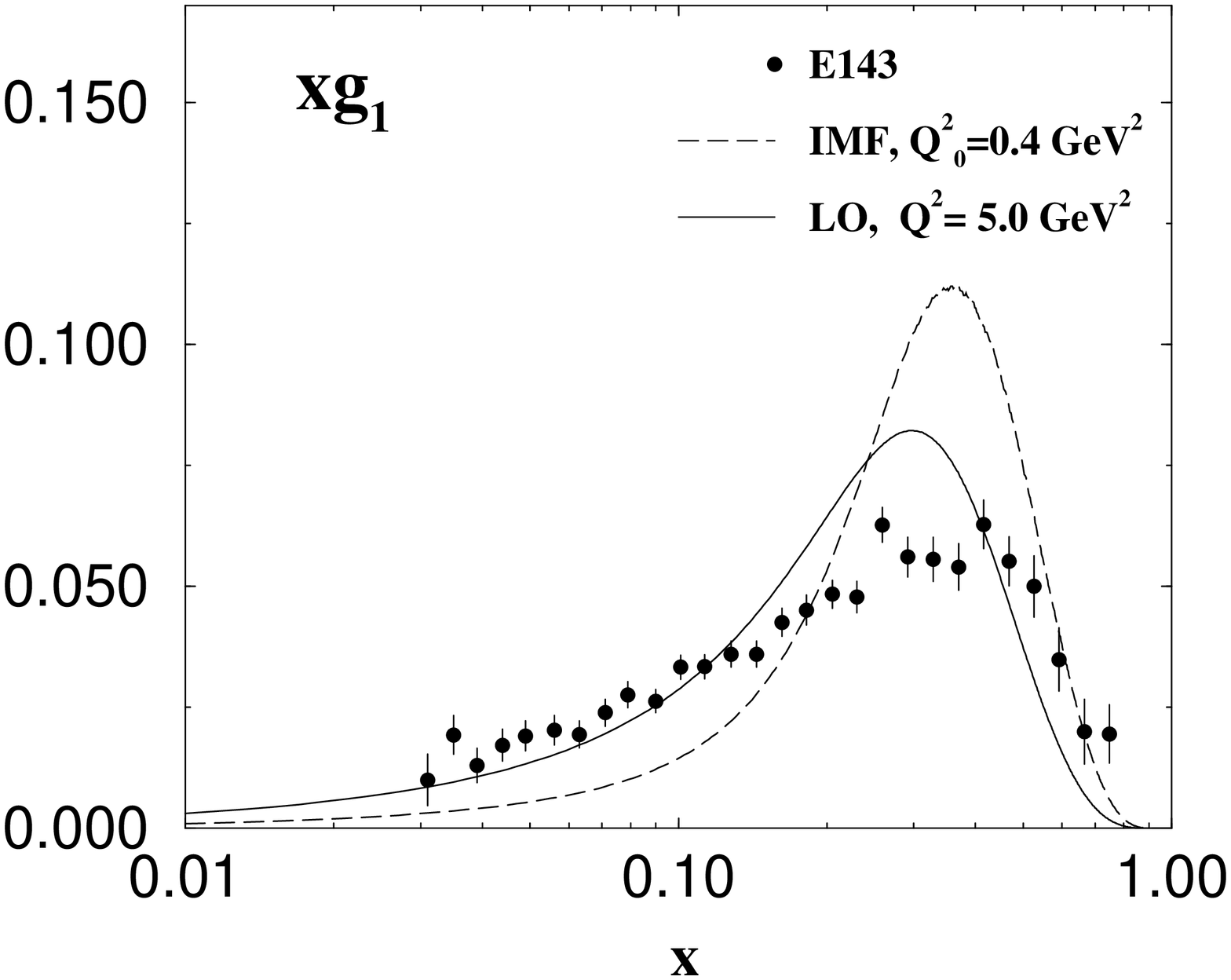,height=6.0cm,width=4.8cm,angle=270}}
\hskip1cm
\parbox[r]{5.2cm}{
\epsfig{figure=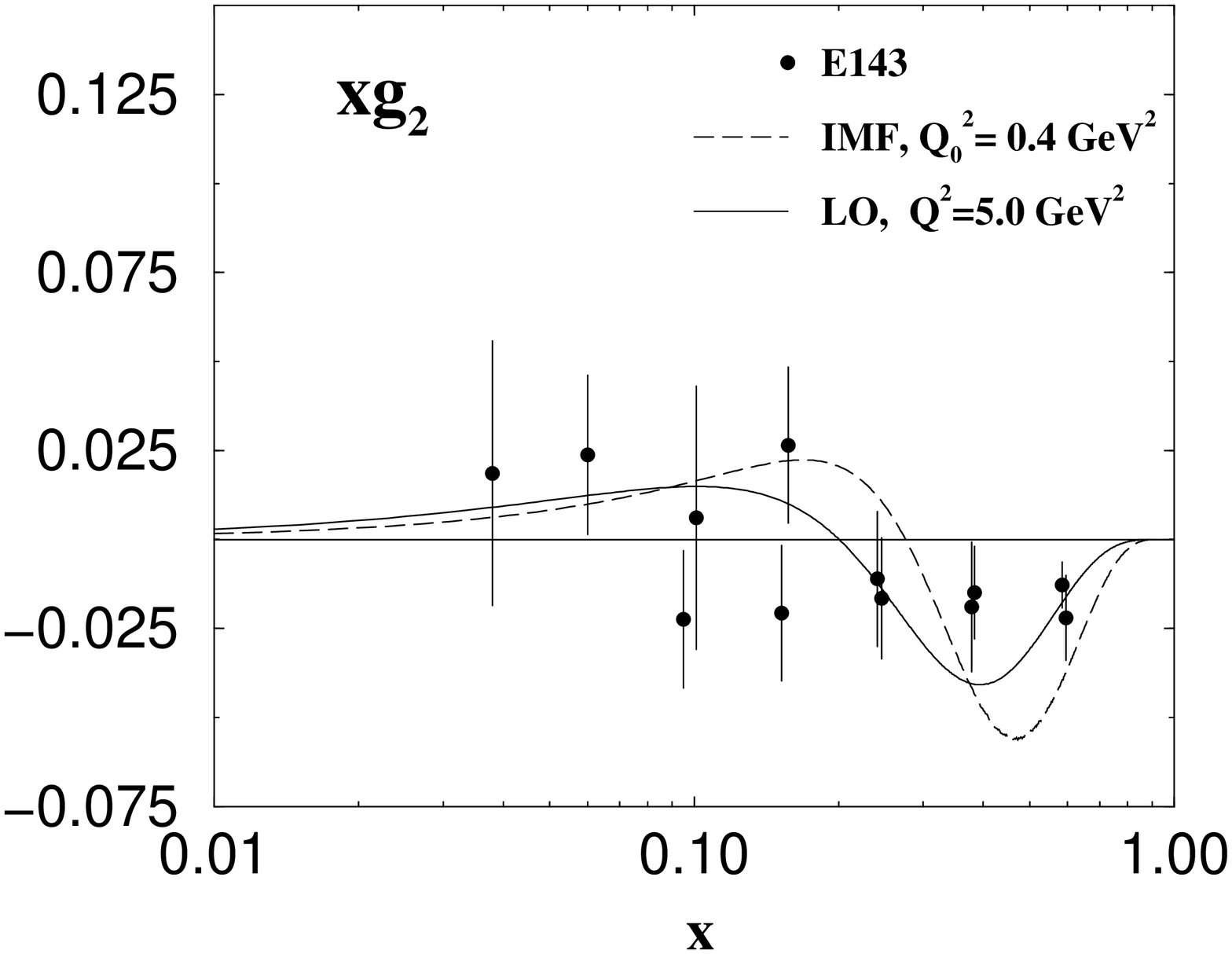,height=6.0cm,width=4.8cm,angle=270}}
\caption{\label{xg12}{\sf Model predictions for the
polarized proton structure functions $xg_1$ (left panel)
and $xg_2$ (right panel). The curves labeled `RF' denote the
results as obtained from the valence quark contribution to
(\protect\ref{simple}). These undergo a projection to the infinite
momentum frame `IMF'~(\protect\ref{IMF}) and a leading order `LO'
DGLAP evolution\protect\cite{DGLAP}. 
Data are from SLAC--E143\protect\cite{Abe98}. }}
\vskip-0.3cm
\end{figure}

\begin{figure}[ht]
\centerline{
\epsfig{figure=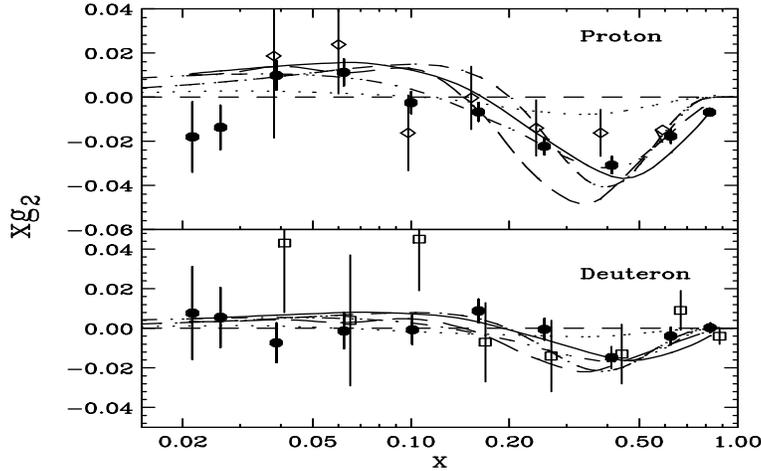,height=6.2cm,width=10.0cm}}
\caption{\label{xg2w}{\sf Model predictions for the
polarized proton structure functions $xg_2$ for proton
and neutron (deuteron) and comparison with data from
E143\protect\cite{Abe98} (open diamond) and 
E155\protect\cite{An02} (open square) and their 
combination (solid circle). The full line is the 
twist--2 truncation~(\protect\ref{g2w}) of data for $g_1(x)$. 
Dashed--dotted\protect\cite{St93} and dotted\protect\cite{So96} 
lines are bag model calculations,
the short dashed lines represent the present chiral soliton 
model\protect\cite{We97} and long dashed line that of 
ref.\protect\cite{Wa00}. (This is a slightly modified figure 
from ref.\protect\cite{An02}.)}}
\vskip-0.3cm
\end{figure}

As in the case for the pion, the model calculation yields the
nucleon structure function at a low energy scale. In addition
the soliton is a localized object. Thus the computed structure 
functions are frame--dependent and one frame has to be picked. 
The appropriate choice is the infinite momentum 
frame (IMF) not only because it makes contact with the parton model 
but also because it is that frame in which the support of the 
structure functions is limited to the physical regime $0\le x\le1$.
Choosing the IMF amounts to the transformation\cite{Ja81,Ga98}
\begin{equation}
f_{\rm IMF}(x)=\frac{1}{1-x}\,f_{\rm RF}(-{\rm ln}(1-x))\, ,
\label{IMF}
\end{equation}
where $f_{\rm RF}(x)$ denotes the structure function as computed
in the nucleon rest frame. So the program is two--stage, first the
transformation of the model structure function to the IMF according 
to eq~(\ref{IMF}) and subsequently the DGLAP evolution program\cite{DGLAP} 
to incorporate the resumed ${\rm ln}Q^2$ corrections. In the current context
it is appropriate to restrain oneself to the leading order (in $\alpha_s$)
in the evolution program because higher orders require the identification 
of quark and antiquark distributions in the parton models sense. In the
present model calculation this is not possible without further 
assumptions\footnote{We assume, however, that the gluon distribution
is zero at the model scale.}. The low energy scale, $Q^2_0=0.4{\rm GeV}^2$, 
at which the model is assumed to approximate QCD has been estimated
in ref.~\cite{We96a} from a best fit to the experimental data
of the unpolarized structure function, $F_2(x)$. The same boundary
value is taken to evolve the model prediction for polarized structure 
function, $g_1(x)$, in the IMF to the scale $Q^2$ of 
several ${\rm GeV}^2$ at which the experimental data are available. 
For the structure function $g_2(x)$ the situation is a bit more 
complicated. First the twist--2 piece must be separated according 
to\cite{Wa77}
\begin{equation}
g_2^{WW}(x)=-g_1(x)+\int_x^1 \frac{dy}{y}\, g_1(y)
\label{g2w}
\end{equation}
and evolved analogously to $g_1(x)$ (which also is twist--2). The
remainder, $g_2(x)-g_2^{WW}(x)$, is twist--3 and is evolved according
to the large--$N_C$ scheme of ref.\cite{Ali91}. Finally, 
the two pieces are again put together at the end--point
of the evolution, $Q^2$. In figure~\ref{xg12} I compare 
the model predictions for the linearly independent polarized structure 
functions of the proton to experimental data\cite{Abe98}.
In figure~\ref{xg2w} I compare the model predictions for both  
the proton and the neutron (in form of the deuteron) not 
only to the recently accumulated data but also to other model
predictions. 
Surprisingly the twist--2 truncation, {\it i.e.}
eq.~(\ref{g2w}) with the data for $g_1(x)$ at the right hand
side, gives the most accurate description of the data. However,
also the chiral soliton model predictions reproduce the
data well. Bag model predictions have a less pronounced
structure.

\begin{figure}[ht]
\begin{center}
\epsfig{figure=a1n.eps,height=5.3cm,width=6.0cm}\hskip0.6cm
\epsfig{figure=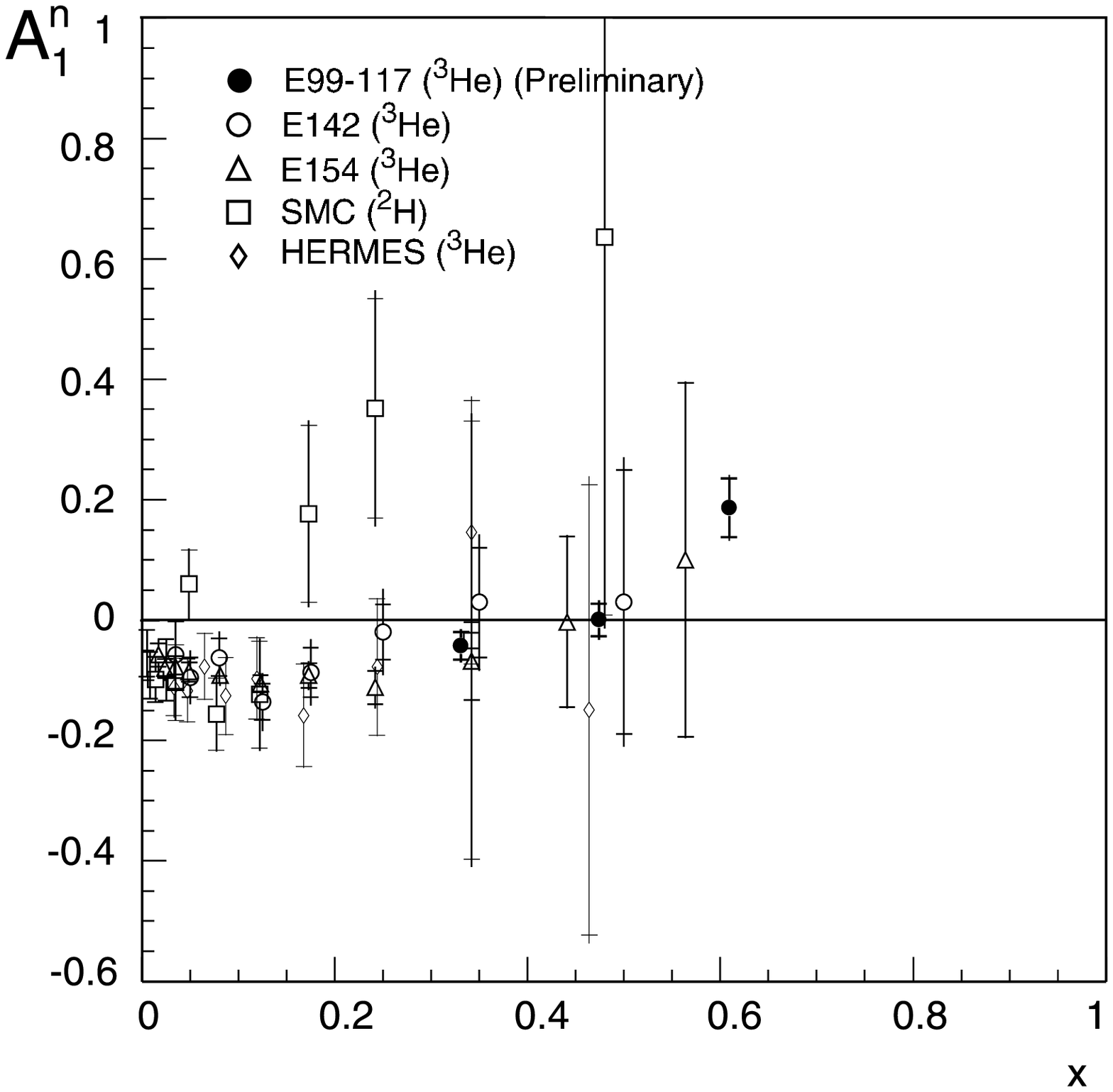,height=5.3cm,width=6.0cm}
\vskip0.2cm
\caption{\label{fig_a1n} Model prediction for $A_1^N$ at different
$Q^2$. The right panel shows the various experimental 
data~\protect\cite{zem}. The JLab--data (E99--117) are still 
preliminary.}
\vskip-0.5cm
\end{center}
\end{figure}

Recently, precise data~\cite{zem} have become available
for the neutron asymmetry
\begin{equation}
A_1=\frac{g_1(x,Q^2)-\frac{4M^2x^2}{Q^2}g_2(x,Q^2)}{F_1(x,Q^2)}\,.
\label{a1n}
\end{equation}
It is therefore challenging to study this quantity in the present
model. As subleading twist contributions are omitted, this amounts
to computing the ratio $g_1(x,Q^2)/F_1(x,Q^2)$, for the neutron.
The resulting ratio is shown in Fig.~\ref{fig_a1n} together with
data. It is interesting
to note that while the ratio at the model scale, $Q_0$, becomes large
and negative at small $x$, the DGLAP evolution causes it to bend around
so that it actually tends to zero as $x\to0$. This behavior is also 
observed from the data, as is the change in sign at moderate $x$.
The position ($x\approx0.25$) at which this change occurs seems 
somewhat lower than the preliminary JLab--data~\cite{zem} suggest and
insensitive to the end point of evolution. Once evolution has set in 
at a moderate point $Q^2$, the evolution to even higher $Q^2$ has 
insignificant effect.

\section{Conclusions}
\vskip-0.1cm
I have discussed a chiral quark model for hadron phenomenology.
In particular, I  considered
the bosonized NJL model as a simplified model for the quark
flavor dynamics.  Although the bosonized version is a meson theory, 
the quark degrees of freedom can indeed be traced. This is very 
helpful for considering structure functions. Additional 
correlations are introduced due to the 
unavoidable regularization which is imposed in a way to respect the
chiral anomaly. Hence a consistent extraction of the nucleon structure 
functions from the Compton amplitude in the Bjorken limit leads to 
expressions that are quite different from those obtained by an 
{\it ad hoc} regularization of quark distributions in the same 
model. I also showed that within a reliable 
approximation the numerical results for the spin dependent 
structure functions agree reasonably well with the empirical 
data.

\vskip-0.3cm
\section*{Acknowledgment}
\vskip-0.1cm
I would like to thank the organizing committee, especially 
Pankaj Jain, for providing this pleasant and worthwhile 
workshop. The contributions of my colleagues R. Alkofer,
L.~Gamberg, H. Reinhardt and E. Ruiz Arriola to
this work are gratefully acknowledged. 
This work has been supported by the Deutsche Froschungsgemeinschaft
(DFG) under contract We 1254/3--2.

\end{document}